\documentclass[10pt,twocolumn,prl,aps,superscriptaddress,floatfix,nobalancelastpage,
preprintnumbers,reprint,nofootinbib,hyperref]{revtex4-2}
\usepackage[colorlinks=true,breaklinks=true]{hyperref}
\usepackage[utf8]{inputenc}
\hypersetup{urlcolor=BlueViolet,
	    citecolor=BlueViolet,
	    linkcolor=BlueViolet}
\usepackage{mathtools}
\usepackage[dvipsnames]{xcolor}

\bibpunct{[}{]}{,}{n}{}{,}

\usepackage{amsmath}
\usepackage{amsfonts}
\usepackage{amssymb}
\usepackage{bbold}
\usepackage{epsfig}
\usepackage{graphicx}
\usepackage{bm}
\usepackage{array}
\usepackage{hyperref}
\usepackage{listings}
\usepackage{color}
\usepackage[normalem]{ulem}
\usepackage{cancel}

\newcommand{\vect}[1]{\boldsymbol{#1}}

\newcommand{\GF}{G_{\rm F}}
\newcommand{\bp}{{\vec{p}}}
\newcommand{\bq}{{\vec{q}}}
\newcommand{\bv}{{\vec{v}}}
\newcommand{\bu}{{\vec{u}}}

\newcommand{\sD}{{\sf D}}
\newcommand{\sS}{{\sf S}}
\newcommand{\sU}{{\sf U}}
\newcommand{\sH}{{\sf H}}
\newcommand{\sM}{{\sf M}}

\newcommand{\vG}{\boldsymbol{G}}
\newcommand{\vR}{\boldsymbol{R}}
\newcommand{\vJ}{\boldsymbol{J}}
\newcommand{\vS}{\boldsymbol{S}}

\newcommand{\vD}{\boldsymbol{D}}

\newcommand{\wP}{\omega_{\rm P}}

\long\def\exclude#1{}

\usepackage{tikz,xcolor,hyperref}

\definecolor{lime}{HTML}{A6CE39}
\DeclareRobustCommand{\orcidicon}{\hspace{-1mm}
	\begin{tikzpicture}
	\draw[lime, fill=lime] (0,0) 
	circle [radius=0.16] 
	node[white] {{\fontfamily{qag}\selectfont \tiny \,ID}};
	\draw[white, fill=white] (-0.0525,0.095) 
	circle [radius=0.007];
	\end{tikzpicture}
	\hspace{-3mm}
}

\foreach \x in {A, ..., Z}{\expandafter\xdef\csname orcid\x\endcsname{\noexpand\href{https://orcid.org/\csname orcidauthor\x\endcsname}
			{\noexpand\orcidicon}}
}

\begin{document}

\preprint{MPP-2021-167}

\title{Neutrino Flavor Pendulum Reloaded: The Case of Fast Pairwise Conversion}

\author{Ian Padilla-Gay\orcidA{}}
\affiliation{Niels Bohr International Academy \& DARK, Niels Bohr Institute,\\ University of Copenhagen, Blegdamsvej 17, 2100 Copenhagen, Denmark}
\author{Irene Tamborra\orcidB{}}
\affiliation{Niels Bohr International Academy \& DARK, Niels Bohr Institute,\\ University of Copenhagen, Blegdamsvej 17, 2100 Copenhagen, Denmark}
\author{Georg G.~Raffelt\orcidC{}}
\affiliation{Max-Planck-Institut f\"ur Physik (Werner-Heisenberg-Institut),\\ F\"ohringer Ring 6, 80805, Munich, Germany}

\date{October 1, 2021}

\begin{abstract}
   In core-collapse supernovae or compact binary merger remnants, neutrino-neutrino refraction can spawn fast pair conversion of the type $\nu_e \bar\nu_e \leftrightarrow \nu_x \bar\nu_x$ (with $x=\mu, \tau$), governed by the angle-dependent density matrices of flavor lepton number. In a homogeneous and axially symmetric two-flavor system, all angle modes evolve coherently, and we show that the nonlinear equations of motion are formally equivalent to those of a gyroscopic pendulum. Within this analogy, our main innovation is to identify the elusive characteristic of the lepton-number angle distribution that determines the depth of conversion with the ``pendulum spin.''  The latter is given by the real part of the eigenfrequency resulting from the linear normal-mode analysis of the neutrino system.  This simple analogy allows one to predict the  depth of flavor conversion without solving the nonlinear evolution equations. Our approach provides a novel diagnostic tool to explore the physics of nonlinear systems.
\end{abstract}

\maketitle


{\em Introduction.}---In neutrino-dense astrophysical environments, such as core-collapse supernovae and the remnants of neutron star mergers, neutrinos experience a significant potential due to the presence of other neutrinos. This refractive effect strongly impacts the flavor evolution of the neutrino radiation field and can lead to collective flavor conversion. While the underlying equations are simple, their nonlinear nature provides for a rich and sometimes confusing plethora of solutions~\cite{Tamborra:2020cul,Chakraborty:2016yeg,Duan:2010bg,Mirizzi:2015eza}.

One case in point is fast pairwise flavor conversion of the type $\nu_e\bar\nu_e\to\nu_x\bar\nu_x$ (where $x$ indicates a generic heavy-lepton flavor, $\mu$ or $\tau$),  conserving the net flavor content and often  called ``fast flavor conversion (FFC).'' Neutrino-neutrino refraction is dimensionally quantified by a typical interaction energy ${\cal O}(\sqrt{2}\GF n_{\nu})$. Specifically, we will use $\mu=\sqrt{2}\GF n_{\nu_e}$ as an overall scale.

Another manifestation of neutrino-neutrino refraction concerns ``slow flavor conversion,'' driven by the energy spectrum and involving flavor exchange between different energy modes. A typical flavor conversion speed is $\sqrt{\omega\mu}$, where $\omega=\Delta m^2/2E$ is the vacuum oscillation frequency depending on the mass-squared difference $\Delta m^2$ and energy $E$. This is defined as ``slow'' because $\mu\gg\omega$. The interpretation of the nonlinear evolution~\cite{Duan:2006an} as a gyroscopic flavor pendulum has been long since 
established~\cite{Hannestad:2006nj,Duan:2007mv,Raffelt:2011yb,Fogli:2007bk,Johns:2017oky} and is the archetype for our study.

Fast flavor conversion is a multi-angle effect of the flavor lepton-number densities. While the nonlinear evolution is a three-flavor problem~\cite{Dasgupta:2007ws,Fogli:2008fj,Dasgupta:2010cd,Friedland:2010sc,Capozzi:2020kge,Shalgar:2021wlj}, the initial instability is between one pair of flavors~\cite{Airen:2018nvp,Banerjee:2011fj,Izaguirre:2016gsx,Capozzi:2019lso}, in practice $\nu_e$ and $\nu_x$. For identical $\nu_x$ and $\bar\nu_x$ distributions, FFC is driven by neutrino electron lepton number (ELN)~\cite{Izaguirre:2016gsx, Capozzi:2017gqd, Morinaga:2021vmc,Yi:2019hrp,Johns:2019izj}, but it is straightforward to include  nontrivial $\nu_x$ and $\bar\nu_x$ distributions~\cite{Shalgar:2021wlj,Capozzi:2020kge,Chakraborty:2019wxe}. An instability of the flavor field requires the ELN angular distribution to change sign at least once---it needs one or more ``crossings.'' In the slow case, spectral crossings are required instead~\cite{Dasgupta:2009mg}. 
 
Various methods have been proposed to identify ELN crossings in hydrodynamical simulations~\cite{Johns:2021taz, Abbar:2020fcl, Dasgupta:2018ulw, Nagakura:2021suv} and significant efforts have been devoted to understand when and where favorable conditions exist for FFC instabilities in astrophysical environments and related implications~\cite{Shalgar:2019kzy, Morinaga:2019wsv, Tamborra:2017ubu, Abbar:2019zoq, Abbar:2018shq, DelfanAzari:2019tez, Nagakura:2019sig, Azari:2019jvr, Capozzi:2020syn,Xiong:2020ntn,Wu:2017qpc, Wu:2017drk,George:2020veu,Abbar:2020qpi,Capozzi:2020syn, Nagakura:2021txn,Nagakura:2021hyb,Johns:2020qsk}. However, despite ELN crossings, only minimal flavor conversion may occur depending on the initial configuration~\cite{Padilla-Gay:2020uxa}, because it is the exact $\nu_e$ and $\bar\nu_e$ angular distribution that determines the ELN crossings and the final flavor outcome~\cite{Johns:2019izj, Shalgar:2019kzy, Shalgar:2019qwg, Padilla-Gay:2020uxa, Yi:2019hrp, Shalgar:2021oko,Bhattacharyya:2021klg}.

This {\it Letter} aims to elucidate under which conditions one should expect large flavor conversion due to FFC in a homogeneous and azimuthally symmetric neutrino gas. For the first time, we  provide a simple diagnostic criterion to predict whether FFC should occur and how much conversion should be expected, without solving the neutrino equations of motion (EOMs). In doing so, we rely on a formal analogy of the EOMs with the ones of a gyroscopic pendulum~\cite{Johns:2019izj}. 

The linear normal-mode analysis~\cite{Airen:2018nvp,Banerjee:2011fj,Izaguirre:2016gsx,Capozzi:2019lso,Sarikas:2012ad,Morinaga:2018aug} has been widely employed to obtain the growth rate of the flavor instability. Our main new insight consists of taking full advantage of this approach and to recognize, for the first time, the fundamental information provided by
the initial rate of precession as well as by the universal form of the linear eigenfunction for the angle-dependent flavor conversion. Such findings provide crucial new insights into the physics of nonlinear systems.


{\em Mean field equations.}---We describe [anti]neutrinos through the usual density matrices $\varrho(\vec{p},\vec{r},t)$  [$\bar\varrho(\vec{p},\vec{r},t)$]. The diagonal elements are occupation numbers, whereas the off-diagonal ones encode flavor coherence. Ignoring collisions, the commutator EOM for neutrinos is~\cite{Raffelt:2007yz}
\begin{equation}\label{eq:eom1}
    i\,(\partial_t+\bv\cdot{\vec\nabla})\varrho_\bp=\left[{\sf\Omega}_E,\varrho_\bp\right]+\sqrt{2}\,\GF\left[\sH_\bv,\varrho_\bp\right]\ ,
\end{equation}
where vacuum oscillations are spawned by ${\sf\Omega}_E=\sM^2/2E$ with $\sM$ being the neutrino mass matrix. Antineutrinos require ${\sf\Omega}_E \rightarrow -{\sf\Omega}_E$, but as we study FFC we set ${\sf \Omega}_E=0$ henceforth, also implying that $\bv=\bp/E$ is a unit vector. The Hamiltonian matrix
\begin{equation}\label{eq:EOM-2a}
  \sH_\bv=\int\!\frac{d^3\bq}{(2\pi)^3}(\varrho_\bq-\bar\varrho_\bq)\,(1-\bv_\bq\cdot\bv)
\end{equation}
represents neutrino-neutrino refraction. The EOMs are understood in a co-moving frame in flavor space such that refraction on ordinary matter disappears.

One central feature of FFC is that all $\varrho(\vec{p},\vec{r},t)$ and $\bar\varrho(\vec{p},\vec{r},t)$, and any linear combination, follow the same EOM that depends on $\bv$ but not on $E$. We thus consider the density matrices for lepton number $\sD_\bp=\varrho_\bp-\bar\varrho_\bp$, which we also integrate over energy and normalize to the $\nu_e$ density. The matrices $\sD_\bv\equiv n_{\nu_e}^{-1}\int_0^\infty {E^2 dE}/{(2\pi^2)}\,\sD_{E,\bv}$ thus defined obey the closed system of equations
\begin{equation}\label{eq:eom2}
    i\,(\partial_t+\bv\cdot{\vec\nabla})\sD_\bv=\mu\left[\sH_\bv,\sD_\bv\right]\ .
\end{equation}
Here, $\mu\equiv\sqrt{2}G_{\rm F} n_{\nu_e}$ is a typical neutrino-neutrino interaction energy, whereas $\sH_\bv=\int (d^2\bu/4\pi)\,D_\bu(1-\bu\cdot\bv)$.

It is perhaps somewhat under-appreciated that it is the energy-integrated lepton-number matrices that drive the entire FFC dynamics. Solving the EOMs amounts to the task of finding $\sH_\bv(t)$. Once it has been found, the solutions for $\varrho_\bp$ and $\bar\varrho_\bp$ or the particle-number densities $\sS_\bp=\varrho_\bp+\bar\varrho_\bp$ can be determined. 

In our case study, we impose several symmetries, the most restrictive one being that of homogeneity of the initial setup {\em and\/} the solutions.  Dropping the gradient and integrating both sides over $\int\!d^2\bv/4\pi$ reveals that the total lepton-number matrix $\sD_0=\int (d^2\bv/4\pi) \sD_{\bv}$ is conserved, meaning that
$n_{\nu_\ell}-n_{\bar\nu_\ell}$ is separately conserved for every flavor $\ell=e$, $\mu$, and $\tau$. Indeed, FFC does not convert any net flavor. 
The corresponding number-density matrix $\sS_0$ is not conserved. While the total particle number (trace of~$\sS_0$) is conserved, the individual $n_{\nu_\ell}+n_{\bar\nu_\ell}$ are not, commensurate with a pair-conversion effect. 

As $\sD_0$ is conserved,  it causes a global precession on the r.h.s.\ of Eq.~\eqref{eq:eom2} that can be ``rotated away'' by the unitary transformation $\sU(t)=\exp[-i\sD_0 t]$ as for the ordinary matter effect. The Hamiltonian matrix  becomes $\sH_\bv=-\bv\cdot\int (d^2\bu/4\pi)\,\bu\,D_\bu$. Note that we have {\em not\/} assumed $\sD_0=0$, we have only absorbed its effect by going to a co-moving frame.
So we recognize that, in the homogeneous case, the evolution is entirely driven by  $\vec{\sD}(t)=\int(d^2\bu/4\pi)\,\bu\,\sD_\bu(t)$. While the latter is not conserved, \smash{${\rm Tr}\,\vec{\sD}^2$} is conserved, meaning that the lepton-number flux, summed over all flavors, is conserved.

As a further simplification, we impose axial symmetry on the initial setup {\em and\/} the solutions. Measuring $\bv$ against the symmetry axis (zenith angle $\theta$), we integrate out the azimuth angle $\phi$ and define $\sD_v=\int_{0}^{2\pi}\! (d\phi/4\pi)\sD_\bv$ where $v=\cos\theta$ is the velocity along the symmetry axis ($v$ is not $|\bv|=1$) with $-1\leq v\leq+1$. The flux matrix now has only one nonvanishing component: \smash{$\sD_1=\int_{-1}^{+1}dv\,v \sD_v$}. A possible factor 1/2 in front of \smash{$\int_{-1}^{+1} d\cos\theta$} has been absorbed in the definition of $\sD_v$.

Last, we consider only two flavors, although three-flavor solutions can be  much richer in the nonlinear regime~\cite{Shalgar:2021wlj, Chakraborty:2019wxe, Capozzi:2020kge, Airen:2018nvp}. The corotating EOM thus becomes
\begin{equation}\label{eq:EOM-vectors}
    i\dot{\sD}_v=\mu v[\sD_v,\sD_1]
    \quad\hbox{or}\quad
    \dot{\vD}_v=\mu\,v\vD_v \times\vD_1\ .    
\end{equation}
We here express the $2\times2$ Hermitian $\sD_v$ matrices through the usual Bloch vectors (polarization vectors) such that $\sD_v=({\rm Tr}\,\sD_v+\vD_v\cdot\vect{\sigma})/2$ with $\vect{\sigma}$ a vector of Pauli matrices. 

The cross product on the r.h.s.\ reveals that the length of each $\vect{D}_v$  is conserved. Moreover, $\vect{D}_0$ and $|\vect{D}_1|$ are conserved. It is  $\vD_1(t)$
that drives the motion of the system and moves like a gyroscopic pendulum~\cite{Johns:2019izj}.


{\em Single-crossed ELN spectra.}---Except for small seeds, our system begins diagonal in the flavor basis where every $\vect{D}_v$ has only a $z$-component. (We use $x$, $y$ and $z$ for directions in flavor space.) So the initial condition is represented by what we call the ELN spectrum,
\begin{equation}
    g_v=D^z_v\big\vert_{t=0}\propto \left(n_{\nu_e}-n_{\bar\nu_e}-n_{\nu_x}+n_{\bar\nu_x}\right)_v\ .
\end{equation}
One or more ``crossings'' ($g_v$ changes sign) are necessary for run-away solutions to exist. This condition is also sufficient for solutions that may break homogeneity and axial symmetry \cite{Morinaga:2021vmc}.

\begin{figure}[ht]
\centering
\includegraphics[width=0.45\textwidth]{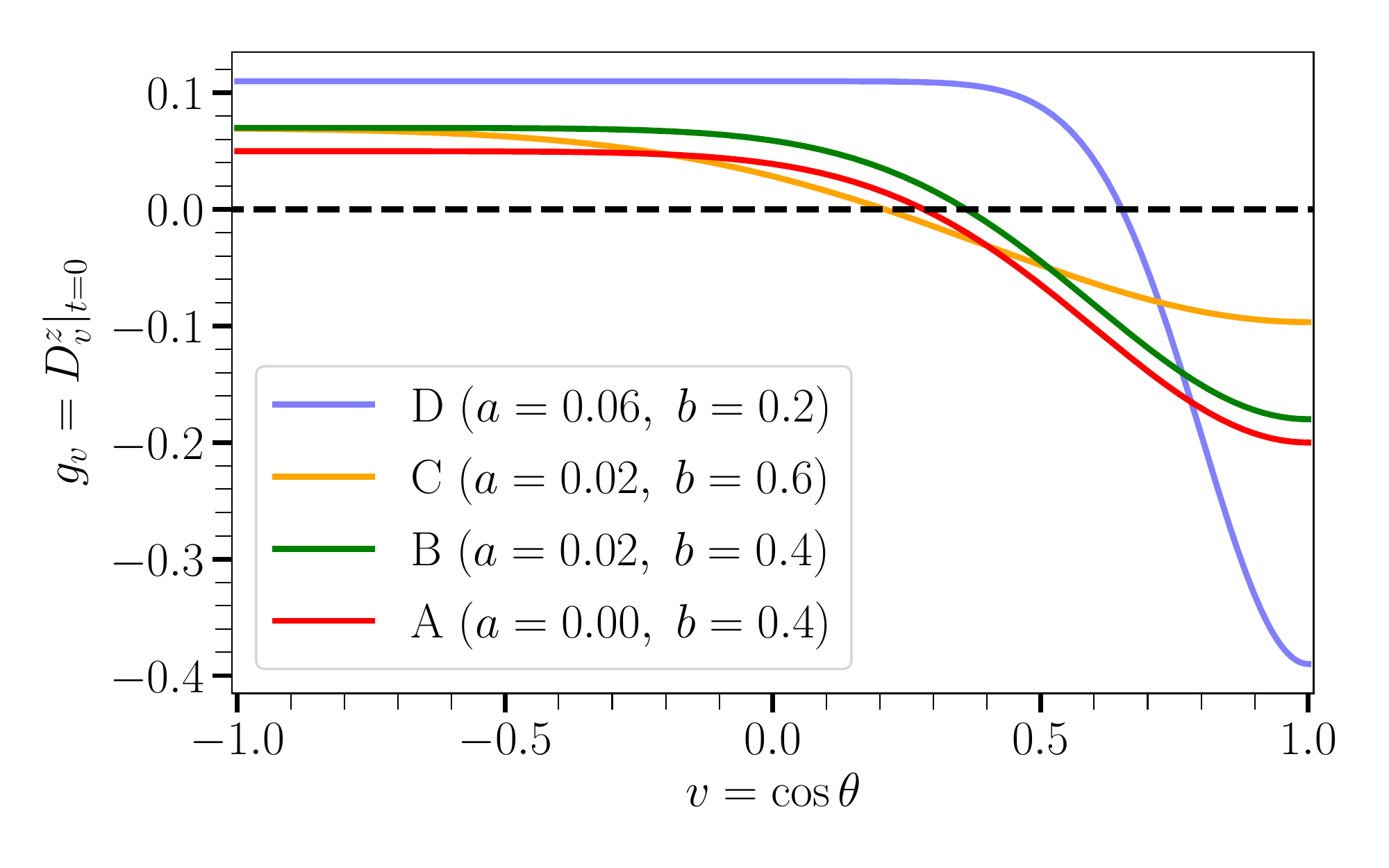}
\vskip-6pt
\caption{Representative ELN distributions $g_v$ defined in  Eqs.~\eqref{eq:init_gaussian_cos} and \eqref{eq:init_gaussian_cos1} for the shown values of $a$ and $b$. \label{fig:1a}}
\vskip15pt
\centering
\includegraphics[width=0.45\textwidth]{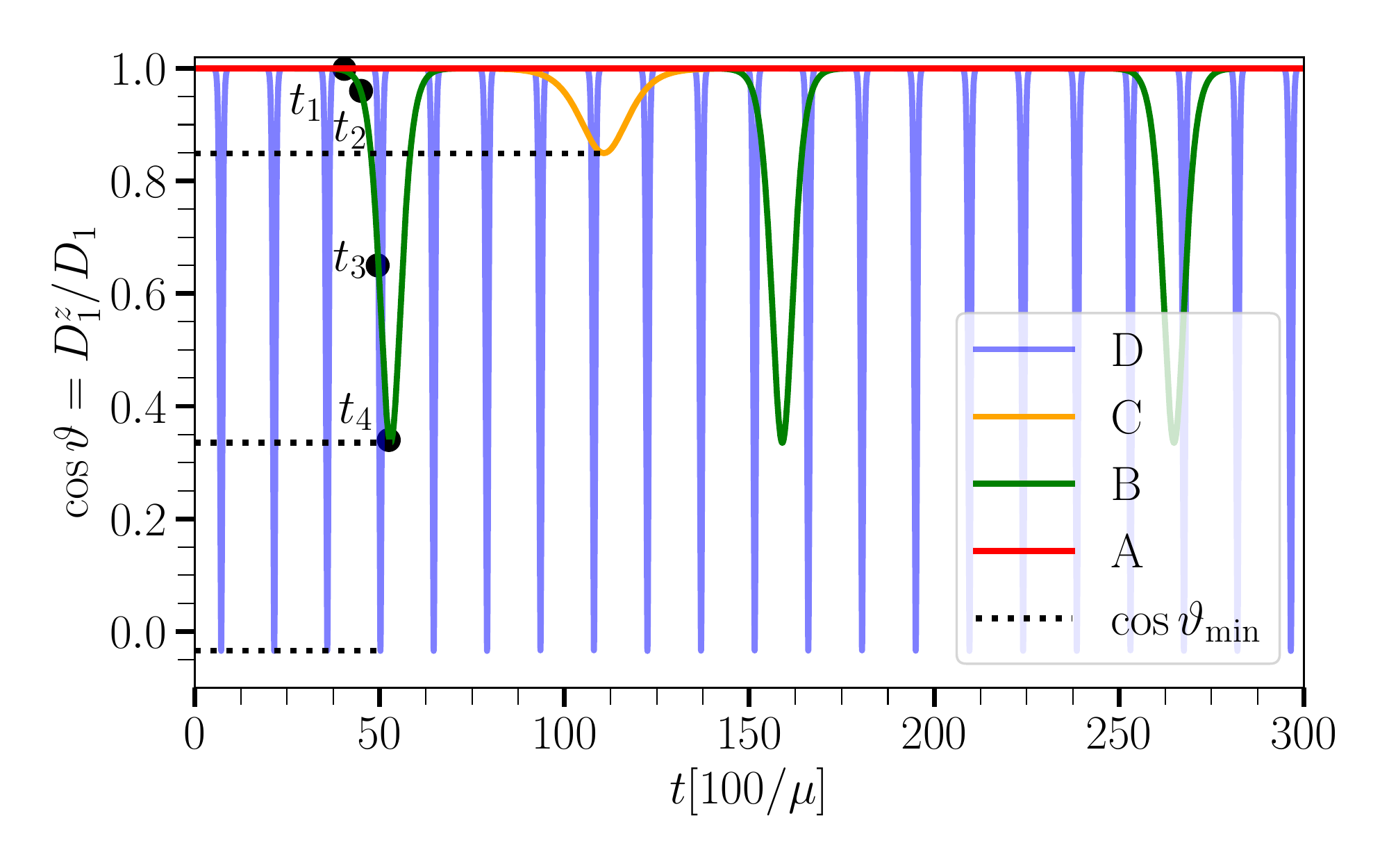}
\vskip-6pt
\caption{Solutions for the $z$-component (flavor direction) of the lepton-number flux $D_1^z(t)$ for the cases A--D specified in Fig.~\ref{fig:1a}, where Case~A has no instability. We show the normalized quantity $\cos\vartheta=D_1^z/D_1$. Its lowest point for each of Cases B--D perfectly agrees  with $\cos\vartheta_{\rm min}$ predicted in Eq.~\eqref{eq:lowest-explicit}.\label{fig:1b}}
\vskip15pt
\centering
\includegraphics[width=0.45\textwidth]{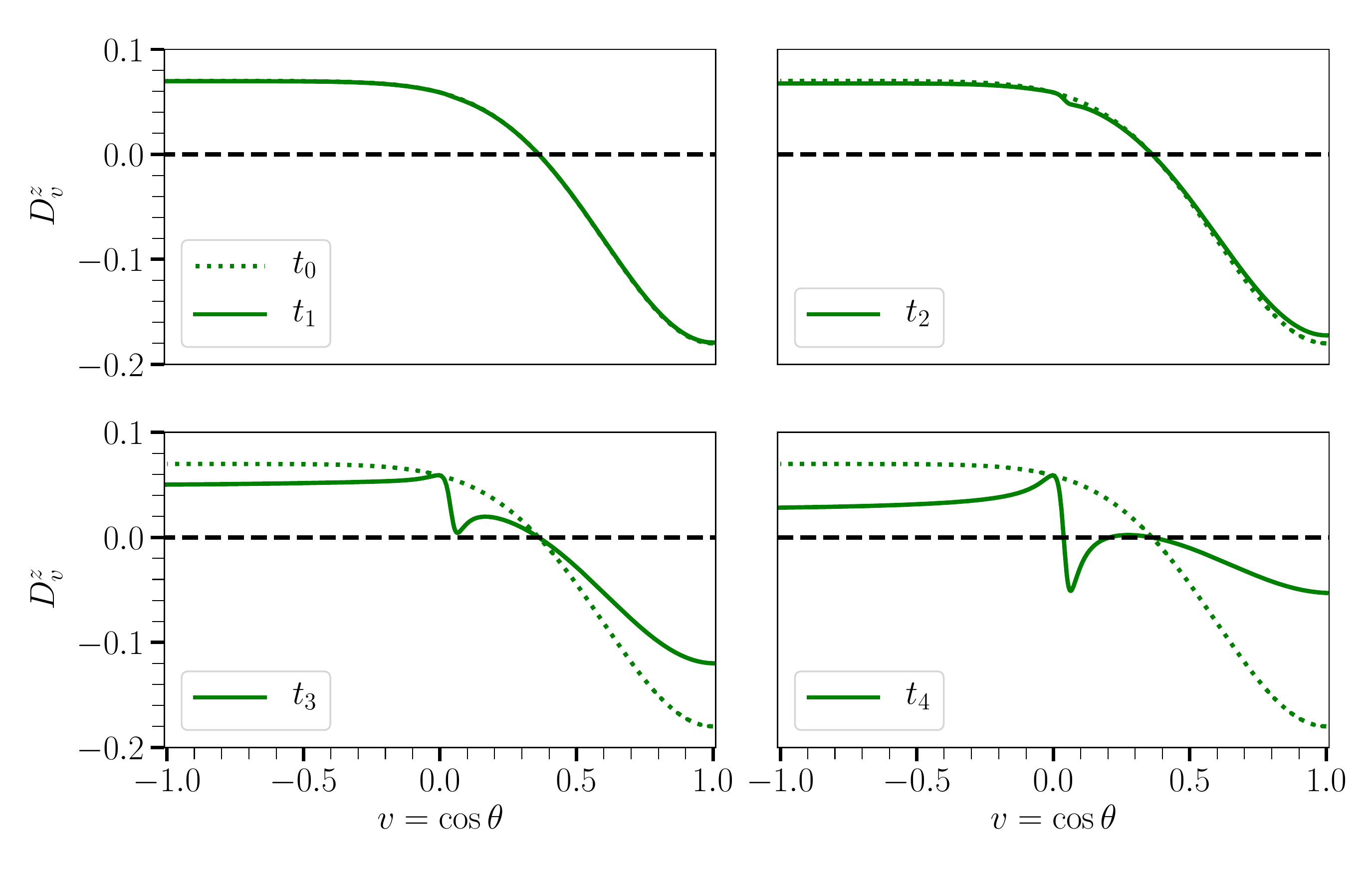}
\vskip-6pt
\caption{Snapshots for $D_v^z(t)$ for case B. The time shots are chosen at $t_1$--$t_4$ indicated in Fig.~\ref{fig:1b} between the beginning of the pendular dip and the maximum excursion. \label{fig:1c}}
\vskip-6pt
\end{figure}

Motivated by the qualitative shape of the ELN angular distributions near the neutrino decoupling regions, we use a family of single-crossed distributions defined by
\begin{subequations}
\begin{eqnarray}
\label{eq:init_gaussian_cos}
\kern-2em\varrho_{ee}(\cos\theta) &=& 0.50\ ,   \\
\kern-2em	\bar{\varrho}_{ee}(\cos\theta)&=&  0.45-a + \frac{0.1}{b}   \exp{\Bigg[\frac{-(1-\cos{\theta})^2}{2 b^2}\Bigg]}\ .
	\label{eq:init_gaussian_cos1}
\end{eqnarray}
\end{subequations}
Here $\int_{-1}^{+1}\varrho_{ee} d\cos{\theta}=1$, whereas the two free parameters $a \in [-0.04,0.12]$ and $b \in [0.1,1]$ determine the shape and normalization of
$g_v=\varrho_{ee}-\bar\varrho_{ee}$ with $v=\cos\theta$. Figure~\ref{fig:1a} shows four representative examples and illustrates the effect of the $a$ and $b$ parameters.

We have solved the EOMs for the cases A--D specified in Fig.~\ref{fig:1a} and show the evolution ${D}_1^z(t)/D_1$ in Fig.~\ref{fig:1b}. Recall that $D_1=|\vD_1|$ is conserved, so we really show $\cos\vartheta$ with $\vartheta(t)$ the zenith angle of $\vD_1(t)$ in flavor space. Case~A has no instability, in agreement with the results of the linear stability analysis, whereas B--D show the characteristic behavior of an inverted pendulum. The waiting time between dips depends logarithmically on the smallness of the chosen seeds. The component $\sqrt{(D_1^{x})^2+(D_1^{y})^2}$ grows exponentially during that period. (For an example, see the Supplemental Material.)


\begin{figure}[ht]
\centering
\includegraphics[width=0.45\textwidth]{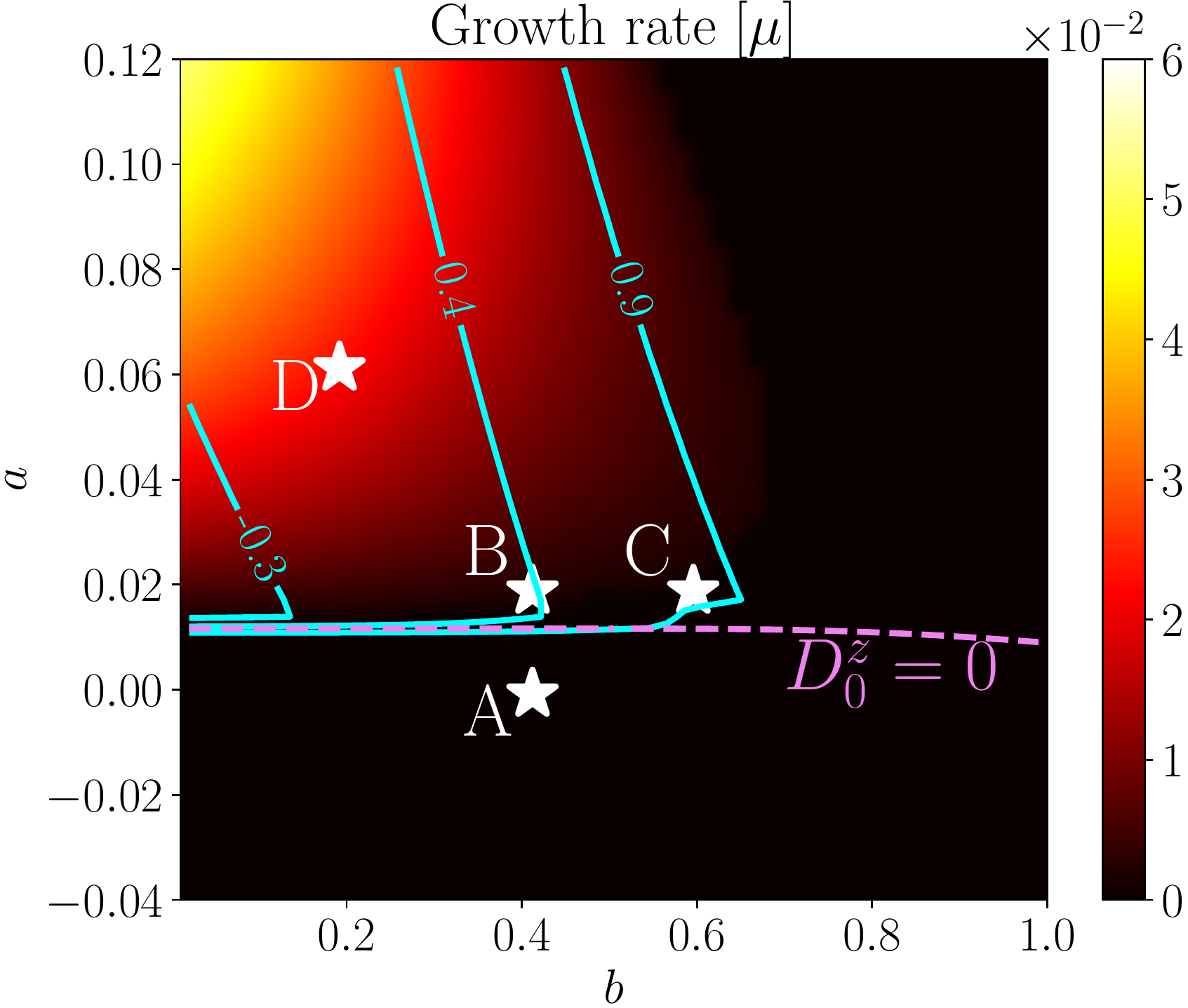}
\caption{Contour plot of the growth rate in the plane spanned by the parameters $a$ and $b$ (see Eqs.~\ref{eq:init_gaussian_cos} and \ref{eq:init_gaussian_cos1}). The white contours represent $D_1^z(t)/D_1|_{\rm min}$. The locus of vanishing lepton number ($D_{0}^z=0$) is marked with a dashed line. We also mark our configurations A--D. We see that large growth rates do not always correspond to large flavor conversion. \label{fig:2a}}
\end{figure}


In Fig.~\ref{fig:1c} we show snapshots of the evolution of the entire spectrum for Case~B at four times indicated in Fig.~\ref{fig:1b}. So we can see how the lepton-number flux  evolves in time as a function of $v=\cos\theta$. All modes evolve coherently and return to their initial position---the overall evolution remains periodic within the limits of numerical precision. The same applies to the analogous evolution of the lepton-number modes $\vect{S}_v(t)$.

Finally, in Fig.~\ref{fig:2a} we show contours of $D_1^z(t)/D_1|_{\rm min}=\cos\vartheta_{\rm min}$ in the plane spanned by $a$ and $b$ overlaid with contours of the growth rate obtained by the linear normal-mode 
analysis~\cite{Banerjee:2011fj,Izaguirre:2016gsx}. Evidently large flavor conversion does not always correlate with a large growth rate. Moreover, seemingly similar ELN configurations can cause very different flavor outcomes.


The coherence of all modes suggests a small number of underlying degrees of freedom. In fact, by applying the Gram matrix method~\cite{Raffelt:2011yb}, we find that   our system with single-crossed ELN spectra is equivalent to  three discrete angle modes, which  form a gyroscopic flavor pendulum in the unstable case (see Supplemental Material for more details).


{\em Pendulum in flavor space.}---The first of the linearly independent functions suggested by the Gram matrix is the conserved vector $\vG=\vD_0=\int dv\,\vD_v(t)$ of lepton number. The second is the lepton-number flux $\vR(t)=\vD_1(t)=\int dv\,v \vD_v(t)$ with conserved length. The third is what we call $\vJ(t)=\int dv\, w_v \vD_v(t)$ with unknown weight function $w_v$. They represent a gyroscopic pendulum, if they obey the EOMs~\cite{Raffelt:2011yb}
\begin{equation}\label{eq:pendulum}
    \dot{\vect{G}}=0\ ,\ 
\dot{\vect{R}}=\mu\vect{J}\times\vect{R} \ \
\mathrm{and}\ \ 
\dot{\vect{J}}=\gamma\vect{G}\times\vect{R}\ .
\end{equation}
In a mechanical analogy, $\vG$ represents gravity, $\vR$ the center-of-mass position relative to the point of support, $\vJ$ the total angular momentum, and $\mu^{-1}$ the moment of inertia. Besides the conserved $\vG$, the EOMs imply four conserved quantities: length $R$ of the radius vector, angular momentum $J_z=\vJ\cdot{\vG}/G$ along ``gravity,'' spin $S=\vJ\cdot\vR/R$, and energy $E=V+T = \gamma\vect{G}\cdot\vect{R}+(\mu/2){\vect J}^2$. Moreover, the natural pendulum frequency $\lambda\mu$ is given by $\lambda^2=\gamma G R/\mu$. We here assume that $\gamma>0$, a possible negative sign is absorbed by redefining $\vG=-\vD_0$.

We use coordinates where $\vG$ defines the $z$-direction so that $\vG=(0,0,G)$, whereas the pendulum is described in polar coordinates $(\vartheta,\varphi)$ by $\vR=R(s_\vartheta c_\varphi,s_\vartheta s_\varphi,c_\vartheta)$ with $s_\vartheta=\sin\vartheta$ and so forth. 

Solving the EOMs for $\vartheta(t)$ and $\varphi(t)$ in terms of the conserved quantities is shown in any mechanics textbook or Appendix~B of Ref.~\cite{Raffelt:2011yb}. One important simplification is that we always begin with $\vR$ parallel or antiparallel to $\vG$ without an initial velocity, implying that $\vJ|_{t=0}=\vS|_{t=0}$, and because $J_z$ and $S$ are conserved, we may use $J_z=S$. Moreover, we assume that $\vS$ is parallel to $\vR$ and not antiparallel. One thus finds
\begin{subequations}
\begin{eqnarray}\label{eq:phi-eq}
   \dot{\varphi}&=&\mu\,\frac{2\lambda \sigma}{1+\cos\vartheta}\ ,
   \\
   \label{eq:theta-eq}
   \dot\vartheta^2&=&\mu^2\lambda^2\Bigl[2\,(1-\cos\vartheta)-\sigma^2\frac{4\,(1-\cos\vartheta)^2}{\sin^2\vartheta}\Bigr]\ ,
\end{eqnarray}
\end{subequations}
where we have expressed the spin, $S=2\lambda\sigma$, in terms of a parameter $\sigma$ and the dimensionless natural pendulum frequency $\lambda=\sqrt{\gamma G R/\mu}$. Besides the overall scale $\mu$, the pendulum is fully described by the parameters $\lambda$ and $\sigma$.

The zenith-angle EOM of Eq.~\eqref{eq:theta-eq} becomes yet more informative with $c_\vartheta=\cos\vartheta$ as independent variable, so that $\dot c_\vartheta^2=\mu^2\lambda^2\,2\bigl(1-c_\vartheta\bigr)^2\bigl(1+c_\vartheta-2\sigma^2\bigr)$. For the r.h.s.\ to be positive in the neighborhood of $c_\vartheta=1$, we obtain $\sigma<1$ as a condition for instability. For larger $\sigma$, the pendulum is stuck in the ``sleeping top position.'' In the unstable case, it nutates between the upright position and a minimal latitude $\vartheta_{\rm min}$ given by 
    $\cos\vartheta_{\rm min}=-1+2 \sigma^2$.
For $\sigma=0$, it reaches the vertical downward position.

In the linear regime ($\vartheta\ll 1$), the solutions~\eqref{eq:phi-eq} and \eqref{eq:theta-eq} are
\begin{equation}\label{eq:pendulum-linear-parameters}
    \dot\varphi=\mu\lambda \sigma \ \
    \mathrm{and} \ \
    \dot\vartheta=\pm\mu\lambda\sqrt{1-\sigma^2}~\vartheta\ .
\end{equation}
The pendulum performs a uniform precession, whereas $\vartheta$ grows or shrinks exponentially, according to whether  the pendulum moves away from the stable position or, after a full swing, comes back to it. 

{\em Normal mode analysis.}---To match these parameters with our full system, we consider the latter in the linear regime. Initially $D_v^{xy}=D_v^x+i D_v^y$ is small, whereas $D_v^z$ is at its initial value $g_v$. Thus the linearized version of  Eq.~\eqref{eq:EOM-vectors} is $(i\partial_t+v D_1)D_v^{xy}=v g_z\int\! du\, u D_u^{xy}$. A collective normal mode would be of the form $g_v Q_v e^{-i\omega t}$ with $\omega=\wP\pm i\Gamma$ being the complex eigenfrequency, where the subscript P stands for ``precession.'' The solution~is
\begin{equation}\label{eq:eigenfunction}
    D_v^{xy}(t)=f\,\frac{v\,g_v}{\omega+v D_1}\,e^{-i\omega t}\ ,
\end{equation}
where $f$ depends on the initial conditions. Inserting this back into the linear EOM reveals that $\omega$ is fixed by
\begin{equation}\label{eq:eigen}
    \int_{-1}^{+1}\!dv\,\frac{v^2 g_v}{\omega+v D_1}=\int_{-1}^{+1}\!dv\,g_vv^2\frac{\wP+vD_1-i\Gamma}{(\wP+v D_1)^2+\Gamma^2}=1\ .
\end{equation}
For convenience, we also provide a step-by-step derivation in the Supplemental Material.

The crucial final step is to match the real and imaginary parts of $\omega$ with the corresponding pendulum parameters of Eq.~\eqref{eq:pendulum-linear-parameters}:   $\dot\varphi = \mu\lambda\sigma =\wP$ and $\dot\vartheta= \pm\mu\lambda\sqrt{1-\sigma^2}~\vartheta = \pm \Gamma \vartheta$. Inverting these relations and selecting the positive solution for the second equation only reveals
\begin{equation}
    \sigma= \sqrt{\frac{\wP^2}{\wP^2+\Gamma^2}}\ \ 
\mathrm{and} \ \
  \lambda=  \frac{1}{\mu}\sqrt{\wP^2+\Gamma^2}\ .
\end{equation}
Hence,  the lowest pendulum position is
\begin{equation}\label{eq:lowest-explicit}
\cos\vartheta_{\rm min}=-1+2\,\frac{\wP^2}{\wP^2+\Gamma^2}\ .
\end{equation}
The equation above crucially links the maximal latitude reached by the gyroscopic pendulum to the real and imaginary parts of the complex eigenfrequency $\omega$, providing a way to predict the depth of flavor mixing without solving the equations of motion. This prediction is in excellent agreement for all our ELN configurations, see  our examples shown in Fig.~\ref{fig:1b} for a comparison. We also see that $\wP=0$ implies $\sigma=0$, leading to complete conversion, whereas $\Gamma=0$ implies $\sigma=1$ and the pendulum is stable. 


{\em Conclusions.}---For a homogeneous and azimuthally symmetric two-flavor neutrino gas, we have explicitly shown that flavor conversion physics strongly depends on details of the ELN distribution. Similar looking angular distributions can lead to completely different outcomes. Notably, the amount of flavor conversion does not directly correlate with the growth rate obtained from the linear normal-mode analysis.

Obvious characteristics are the conserved Bloch vector of the lepton number that we call $\vD_0$ and the one of lepton-number flux $\vD_1$ with conserved length, and it is also evident that $\vD_1(t)$ is what drives the evolution of all Bloch vectors (or density matrices) for individual modes of lepton or particle number. 

The evolution of $\vD_1(t)$ appears to be equivalent to a gyroscopic pendulum, with $\vD_0$ playing the role of gravity, suggesting that the third characteristic is what plays the role of spin or equivalently the total angular momentum $\vJ$. However, identifying $\vJ$ as a simple combination of $\vect{D}_n=\int dv\,v^n \vD_v$ is not generally successful~\cite{Johns:2019izj}.

Our main innovation was to match the pendulum parameters (natural frequency and spin) with the precession frequency $\wP$ and growth rate $\Gamma$ obtained from the usual normal-mode analysis of the neutrino system. It is important to stress that, while attention was usually focused on $\Gamma$, the previously ignored $\wP$ provides the spin and thus allows one to gain insight on the amount of flavor mixing.

Our work provides new insights and a simple tool to unveil the rich phenomenology of FFC, shedding light on the complicated physics of neutrino-dense media and, in general, nonlinear systems of this type. While our findings are based on a single-crossed, homogeneous and azimuthally symmetric neutrino gas, they could provide a first step to analytically forecast the amount of flavor conversion in neutrino-dense astrophysical environments. As such, this work could have fundamental implications on our understanding of neutrino flavor evolution in core-collapse supernovae and the synthesis of heavy elements in compact binary merger remnants, where progress is currently halted by its intrinsic numerical challenges.


{\em Acknowledgments.}---We thank Lucas Johns for helpful comments on our manuscript. We are grateful to the Villum Foundation (Projects Nos.~13164 and 37358), the Danmarks Frie Forskningsfonds (Project No.~8049-00038B), and the Deutsche For\-schungs\-ge\-mein\-schaft
through Sonderforschungbereich SFB~1258 ``Neutrinos and Dark Matter in Astro- and Particle Physics.'' 


\bibliographystyle{bibi}
\bibliography{references}

\onecolumngrid
\appendix

\clearpage

\setcounter{equation}{0}
\setcounter{figure}{0}
\setcounter{table}{0}
\setcounter{page}{1}
\makeatletter
\renewcommand{\theequation}{S\arabic{equation}}
\renewcommand{\thefigure}{S\arabic{figure}}
\renewcommand{\thepage}{S\arabic{page}}

\begin{center}
\textbf{\large Supplemental Material\\[0.5ex]
Neutrino flavor pendulum reloaded: The case of fast pairwise conversion}
\end{center}

In this Supplemental Material, we introduce the multipole decomposition of the EOM and show that the pendulum equations derived by truncating the multipole equations to the first few multipoles are not always predictive of the final flavor outcome. Next, we  perform a discretization of the ELN angular distribution to  three  modes and derive a formal similarity with a pendulum characterized by its natural frequency and spin. We also outline the linear normal-mode analysis for our homogeneous system and finally provide additional details on our numerical examples.  

\bigskip

\twocolumngrid

\section{A.~Multipole decomposition} 

One way to discretize the system of interacting Bloch vectors $\vD_v$ with $v=\cos\theta$ is an expansion in Legendre polynomials $L_l(v)$ that more generally would appear in a multipole expansion of $\vD_\bv$ before assuming azimuthal symmetry~\cite{Raffelt:2007yz}. Thus, we define the new functions
\begin{equation}
    \vD_l(t)=\int_{-1}^{+1}dv\,L_l(v)\,\vD_v(t)
\end{equation}
that obey the co-rotating EOMs
\begin{equation}\label{eq:SD}
    	\dot{\vect{D}}_{l} = \frac{\mu}{2} \bigl( a_{l}\vect{D}_{l-1} + b_{l}\vect{D}_{l+1} \bigr)\times\vect{D}_1\ ,
\end{equation}
where $a_{l} = 2l/(2l+1)$ and $b_{l} = 2(l+1)/(2l+1)$. The EOMs for the first few multipoles are explicitly:
\begin{subequations}\label{eq:LowMultipoles}
\begin{eqnarray}
\dot{\vect{D}}_0&=&0\ ,
\\[1ex]
\dot{\vect{D}}_1&=&\mu\,\frac{\vect{D}_0+2\vect{D}_2}{3}\times\vect{D}_1\ ,
\\[1ex]
\dot{\vect{D}}_2&=&\frac{3\mu}{5}\,\vect{D}_3\times\vect{D}_1\ ,
\\[1ex]
\dot{\vect{D}}_3&=&\mu\,\frac{3\vect{D}_2+4\vect{D}_4}{7}\times\vect{D}_1\ .
\end{eqnarray}
\end{subequations}
$\vect{D}_0$ is conserved and $\vect{D}_1$ is the only one that evolves instantaneously like a precession, i.e., its length is conserved. The equation for $\vect{D}_3$ is the first one clearly showing the general structure that a given $\vD_l$ is driven by one higher and one lower multipole.

This infinite tower of equations can be closed by truncation, assuming that the spectrum has no fine-grained information. In this case, high multipoles should be considered to be negligible. Actually, this is a nontrivial point because  it looks like lower multipoles impact higher ones in the EOMs, so higher multipoles should get excited from lower ones, even if they were small at first, as also discussed in Refs.~\cite{Raffelt:2007yz,Johns:2020qsk}.

Johns et al.~\cite{Johns:2019izj} have observed that, if we truncate Eqs.~\eqref{eq:SD} by setting $\dot{\vect{D}}_3=0$, the lowest multipole equations in the comoving frame are equivalent to the ones of a pendulum in the flavor space. Comparing Eqs.~\eqref{eq:LowMultipoles} with Eqs.~\eqref{eq:pendulum} reveals that we should identify $\vR=\vD_1$ as usual and $\vJ=(\vD_0+2\vD_2)/3$, implying $\dot\vJ=2\dot\vD_2/3$. In turn, this implies that we may identify $\vG= 2\vD_3/5$ and $\gamma=\mu$. We now denote with $D_n=D_n^z|_{t=0}$ the initial values that are not conserved except for $\vD_0$ and $\vD_1$. With this notation, one finds for the spin $S=(D_0+2D_2)/3$ and finally
\begin{subequations}
\begin{eqnarray}
   \lambda^2&=&\frac{2}{5}\, D_3D_1\ ,
   \\[1ex]
   \sigma^2&=&\frac{S^2}{4\lambda^2}=\frac{(1/9)\,(D_0+2D_2)^2}{(8/5)\,D_1 D_3}\ .
\end{eqnarray}
\end{subequations}
With these identifications, our interpretations agree with the ones in Ref.~\cite{Johns:2019izj}, noting that they used the symbol $\sigma$ for what we call $S$. Hence, the condition for an instability $\sigma<1$ reads
\begin{equation}\label{eq:sigma1}
    \frac{(D_0+2D_2)^2}{D_1D_3}<\frac{72}{5}\ .
\end{equation}
Or, equivalently, the pendulum is locked in its initial configuration if
\begin{eqnarray}
\label{eq:xi}
	\xi = \frac{S^2}{(2/5)\,D_{1}^z D_{3}^z} > 4\ .
\end{eqnarray}

From the relations $\Gamma=\mu\lambda\sqrt{1-\sigma^2}$ and $\cos\vartheta_{\rm min}=-1+2\sigma^2$ provided in the main text, these results imply
\begin{subequations}
\begin{eqnarray}
   \Gamma&=&\mu\sqrt{\frac{2D_1D_3}{5}-\frac{(D_0+2D_2)^2}{36}}\ ,
   \\[1ex]
   \cos\vartheta_{\rm min}&=&-1+\frac{5\,(D_0+2D_2)^2}{36\,D_1D_3}
\end{eqnarray}
\end{subequations}
for the initial growth rate and depth of conversion. These predictions can be compared with those of our numerical examples, or equivalently, with those from the normal-mode analysis. 

Actually, as a starting point for their pendulum discussion, the authors of Ref.~\cite{Johns:2019izj} used the second-order equation
\begin{equation}
\label{eq:pendulum_app}
	\frac{\vect{r} \times \vect{\ddot{r}}}{\mu} + S \vect{\dot{r}} = \mu D_{1} \vect{G} \times \vect{r}\ ,
\end{equation}
where $\vect{R}=\vect{D}_1$, $\vect{r} = \vect{R}/R$, the spin of the pendulum is $S = \vect{r}  \cdot (\frac{1}{3}\vect{D}_{0}+\frac{2}{3}\vect{D}_{2})$, and $\vect{G} = \frac{2}{5}\vect{D}_{3}$. 

To show that this second-order equation follows from our two first-order ones, we write the latter in the form $\dot{\vect{r}}=\vJ\times\vect{r}$ and $\dot\vJ=\vG\times\vect{r}$ where $\mu$ was absorbed in the definition of time and all other coefficients in the definition of $\vG$. Taking another derivative of the first equation yields $\ddot{\vect{r}}=\dot\vJ\times\vect{r}+\vJ\times\dot{\vect{r}}$ and inserting the second equation for $\dot\vJ$ gives $\ddot{\vect{r}}=(\vG\times\vect{r})\times\vect{r}+\vJ\times\dot{\vect{r}}=
(\vG\cdot\vect{r})\vect{r}-\vG+\vJ\times\dot{\vect{r}}$, where we have used $\vect{r}^2=1$. Now we consider $\vect{r}\times\ddot{\vect{r}}$ and see that the first term disappears and the second is $\vG\times\vect{r}$; the third is $\vect{r}\times(\vJ\times\dot{\vect{r}})=(\vect{r}\cdot\dot{\vect{r}})\vJ-(\vect{r}\cdot\vJ)\dot{\vect{r}}$.
Noting that $\dot{\vect{r}}$ is perpendicular to $\vect{r}$ and  $\vect{r}\cdot\vJ=S$ is the conserved spin,   we find
$\vect{r}\times\ddot{\vect{r}}+S\,\dot{\vect{r}}=\vG\times\vect{r}$. Reinstating the original meaning of the variables leads to Eq.~\eqref{eq:pendulum_app}. The advantage is that $\vJ$ no longer appears, but only the conserved spin.

As already discussed in the main text, for sufficiently large $S$, the pendulum is locked in its initial stable configuration and cannot swing away from the flavor axis. Hence, the pendulum is stable, and we expect that FFC cannot take place. The pendulum is also in a stable configuration when it is oriented in the same direction as  the gravity vector $\vect{G}$. 

\begin{figure}[b!]
\centering
\includegraphics[width=\columnwidth]{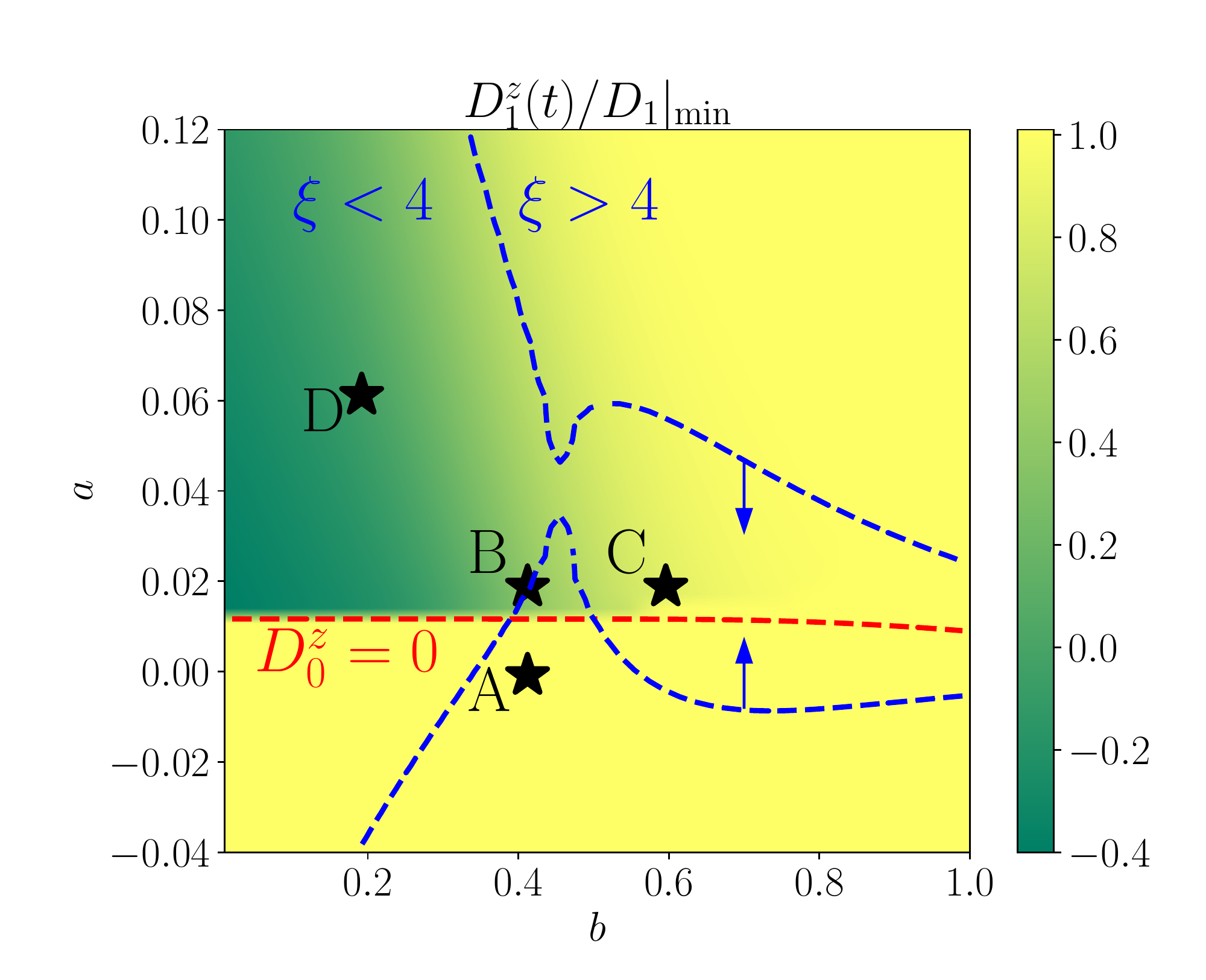}
\caption{Contour plot of the minimum value of the lepton-number flux $D_1^z(t)/D_1^z|_{t=0}$  in the plane spanned by $a$ and $b$ (see Eqs.~\ref{eq:init_gaussian_cos}, \ref{eq:init_gaussian_cos1} and Fig.~\ref{fig:2a}).    The  isocontour of the configurations with  $\xi=4$ (obtained by truncating the multipole expansion of the EOM at $l=3$) is marked by  dashed blue lines, where the arrows point into the unstable region ($\xi \lesssim 4$). Due to the limited number of multipoles, this criterion worsens for very peaked angular distributions and does not allow one to reliably predict the flavor outcome for a general ELN configuration. }
\label{fig:1appendix}
\end{figure}
Figure~\ref{fig:1appendix} shows the contour plot of the minimum value of the lepton-number flux $D_1^z(t)$   in the plane spanned by $a$ and $b$. We can see from Fig.~\ref{fig:1appendix} (see also  Fig.~\ref{fig:2a}) that we expect a different flavor outcome as a function of $a$ and $b$, with regions of no flavor mixing despite the existence of an ELN crossing. 

In agreement with Ref.~\cite{Johns:2019izj}, our results support that the outcome of the neutrino flavor qualitatively changes depending on the relative signs of the $l=0$--$3$ multipoles. The relative sign of $\vect{D}_{0}|_{t=0}$ and $\vect{D}_{2}|_{t=0}$ determines the magnitude of $S$. Secondly, the relative sign of $\vect{D}_{1}|_{t=0}$ and $\vect{D}_{3}|_{t=0}$ also determines whether the pendulum is initially in a stable (unstable) equilibrium configuration, before it is perturbed. Our results are also in agreement with the ones of Ref.~\cite{Johns:2019izj} for the configurations when gravity plays a role in stabilizing the pendulum (results not shown here). Qualitatively, these findings are in agreement with Fig.~3 of Ref.~\cite{Johns:2019izj} where the relative signs of the multipoles lead to different growth rates of the flavor instability. 

This suggests that it may be enough to rely on the $l=0$--$3$ multipoles of $\vect{D}_v$ in order to predict the stability of the flavor pendulum and gauge the amount of flavor mixing. However, it is important to stress  that the errors induced by truncating the angular-moment expansion at an arbitrary small $l$ propagate back to large scale with major consequences on the overall flavor evolution in the nonlinear regime~\cite{Johns:2020qsk}.

The $\xi=4$ isocontour (dashed blue line) in Fig.~\ref{fig:1appendix} shows unstable regions predicted by the pendulum analysis. We find that in the stable (bottom) part of the parameter space in Fig.~\ref{fig:1appendix}, the spin is large enough to lock the flavor pendulum, not letting it swing away from the flavor axis.  Conversely, in the unstable (top) region in Fig.~\ref{fig:1appendix}, $S$ is smaller, allowing $\vect{r}$ to oscillate. However, we also find that fast flavor mixing does not occur for  all configurations below the black dashed line representing the locus where $D_{0}^z=0$ and the unstable regions do not coincide with the $\xi = 4$ contour.

This discrepancy is due to the fact that, for very forward-peaked distributions, the $\xi$ criterion worsens, see the lower region below the $D_{0}^z=0$ line. Moreover, we can see a sudden transition to large flavor mixing in the proximity of the $D_{0}^z=0$ line, not predictable by the $\xi$ criterion. These deviations of the numerical results from the $\xi=4$ constraints are due to the fact  the high-$l$ multipoles (with $l>3$) are relevant and do affect the flavor stability. As a consequence, it is difficult to asses, a priori, when the pendulum approximation proposed in Ref.~\cite{Johns:2020qsk} should hold.

\section{B. Flavor pendulum of three modes}
The coherence of all modes suggests a small number of underlying degrees of freedom that can be diagnosed using the Gram matrix $G_{ij}=\int_{t_1}^{t_2}dt\,\vect{D}_{v_i}(t)\cdot\vect{D}_{v_j}(t)$ \cite{Raffelt:2011yb}. It is calculated for our discrete set of numerical $\vect{D}_{v_i}(t)$ with $i=1,\ldots,n$ for a convenient, but arbitrary, time interval. The rank of $G$ that we call $N+1$ reveals the number of independent functions. The system always has one time-independent solution in the form of $\vD_0=\sum_{i=1}^n\vect{D}_{v_i}(t)$, thus $N$ is the number of independent {\em dynamical} functions. For our single-crossed examples we always find $N=2$. Hence, we conjecture that single-crossed ELN spectra provide solutions that are equivalent to two dynamical degrees of freedom, equivalent to three discrete angle modes.

To study a system of three discrete modes we note that another way to combine the $\vect{D}_v$ is to use angular moments of the $v=\cos\theta$ distribution defined as
\begin{equation}
    \vect{M}_n=\int_{-1}^{+1}\! dv\,v^n\vect{D}_v\ .
\end{equation}
Here $\vect{M}_0$ is the same as $\vect{D}_0$ and $\vect{M}_1=\vect{D}_1$ is the flux. The EOM is
\begin{equation}
 \dot{\vect{M}}_n = \mu\,\vect{M}_{n+1}\times \vect{M}_{1}\ .
\end{equation}
Once more we see immediately that $\vect{M}_0$ is conserved, whereas the dipole $\vect{M}_1$ performs an instantaneous precession around the second moment and thus its length $M_1=|\vect{M}_1|$ is conserved. The length of the other moments is not conserved. The Legendre polynomials (see Appendix~A) are one combination of the moments that is based on an orthogonal set of functions, whereas the $v^n$ are linearly independent, but not orthogonal.

The EOMs should be discretized to be solved numerically. We will see that the evolution is coherent among the $\vect{D}_v$, meaning that neighboring modes have similar evolution and do not develop large differences over time. In this sense, representing the spectrum with a small discrete set of modes  $v_i$ with $i=1,\ldots,N$ should provide a good proxy to the true solution. Moreover, in our axially symmetric case, there are no spurious instabilities~\cite{Sarikas:2012ad,Morinaga:2018aug}.

Notice, however, that we need a minimum of three discrete bins (or ``beams'') to obtain nontrivial results. As in the continuous case, the overall lepton number $\vect{D}_0=\sum_{i=1}^N\vect{D}_{v_i}$ is conserved and $\vect{D}_1=\sum_{i=1}^N v_i\vect{D}_{v_i}$ has conserved length. So if $N=2$ the only possible solution is a precession of $\vect{D}_1$ around $\vect{D}_0$. For $N\geq 3$, there exist instabilities and pendulum-like solutions.

Next we consider the simplest homogeneous case that can provide an instability, i.e., the general three-mode case consisting of three Bloch vectors $\vect{D}_{v_i}$ with velocities $v_i$ with $i=1$, 2 or~3. The corresponding angular moments are $\vect{M}_n=\sum_{i=1}^3 v_i^n \vect{D}_{v_i}$. In turn, we can express the three $\vect{D}_{v_i}$ in terms of the moments. We have only three $\vect{D}_{v_i}$ modes, so there are only three linearly independent moments. 

We can express any moment in terms of three others. We use the lowest ones and close the tower of EOMs with
\begin{eqnarray}
\vect{M}_3&=&v_1v_2v_3\vect{M}_0+(v_1+v_2+v_3)\vect{M}_2
\nonumber\\
&&\kern4em{}-(v_1v_2+v_1v_3+v_2v_3)\vect{M}_1\ .
\end{eqnarray}
To find this result, we first expressed the three $\vD_{v_i}$ in terms of the first three moments, and then inserted these expressions in the definition of $\vect{M}_3$. Then the tower of EOMs for the moments is
\begin{subequations}
\begin{eqnarray}
\kern-2em\dot{\vect{M}}_0&=&0\ ,
\\
\kern-2em\dot{\vect{M}}_1&=&\mu\vect{M}_2\times\vect{M}_1\ ,
\\
\kern-2em\dot{\vect{M}}_2&=&\mu\vect{M}_3\times\vect{M}_1
\nonumber\\
\kern-2em&=&\mu\bigl[v_1v_2v_3\vect{M}_0 +(v_1+v_2+v_3)\vect{M}_2\bigr]\times\vect{M}_1\,.
\end{eqnarray}
\end{subequations}
We see that we can add any multiple of $\vect{M}_1$ to $\vect{M}_2$ without changing the second equation. Specifically we use
\begin{eqnarray}
\vect{M}^\prime_2&=&\vect{M_2}-(v_1+v_2+v_3)\vect{M}_1
\nonumber\\
&=&-(v_2+v_3)v_1\vect{D}_{v_1}-(v_1+v_3)v_2\vect{D}_{v_2}
\nonumber\\
&&\kern7.4em{}-(v_1+v_2)v_3\vect{D}_{v_3}\ ,
\end{eqnarray}
providing the EOMs
\begin{subequations}
\begin{eqnarray}
\dot{\vect{M}}_0&=&0\ ,
\\
\dot{\vect{M}}_1&=&\mu\vect{M}'_2\times\vect{M}_1\ ,
\\
\dot{\vect{M}}^\prime_2&=&\mu\,v_1v_2v_3\vect{M}_0\times\vect{M}_1\ .
\end{eqnarray}
\end{subequations}
These are the pendulum equations in the form of Eq.~\eqref{eq:pendulum} with the identification $\vect{G}=\vect{M}_0$ (gravity), $\vect{R}=\vect{M}_1$ (pendulum radius), $\vect{J}=\vect{M}_2'$ (angular momentum), and the coupling constant $\gamma=\mu v_1v_2v_3$. If $v_1v_2v_3$ of the chosen beams is negative, we instead identify $\vG=-\vect{M}_0$ to ensure a positive $\gamma$.

These results imply $\lambda^2=v_1v_2v_3 M_0 M_1$ for the square of the natural pendulum frequency, whereas the spin is
$S=J_z=M_2-(v_1+v_2+v_3)M_1$, where we use $M_2=M_2^z|_{t=0}$, recalling that the length of $\vect{M}_2$ is not conserved. The condition for instability is $S^2<4\lambda^2$ or explicitly
\begin{equation}
    \bigl[M_2-(v_1+v_2+v_3)M_1\bigr]^2 < 4 |v_1v_2v_3 M_0 M_1|\ .
\end{equation}
So none of $v_i$ must vanish and, of course, the lepton number $M_0$ and lepton-number flux $M_1$ both must be nonzero. To have three modes in the first place, all three $D_{v_i}^z$ must be nonzero, so all six parameters of our model must be nonzero. The quantity representing the angular momentum is complicated and does not suggest any simple extension to a continuous spectrum.

For given pendulum parameters we can find an equivalent three-mode system. The reverse transformation applied to the initial configuration provides
\begin{subequations}
\begin{eqnarray}
g_{v_1}&=&\frac{S+v_1M_1+v_2v_3 M_0}{(v_1-v_2)(v_1-v_3)}\ ,
\\
g_{v_2}&=&\frac{S+v_2M_1+v_1v_3 M_0}{(v_2-v_1)(v_2-v_3)}\ ,
\\
g_{v_3}&=&\frac{S+v_3M_1+v_1v_2 M_0}{(v_3-v_1)(v_3-v_2)}\ ,
\end{eqnarray}
\end{subequations}
where we have used $J_z=S$ and the spectrum of discrete modes is $g_{v_i}=D_{v_i}^z|_{t=0}$.

As discussed in the main text, from a single-crossed spectrum $g_v$ we can obtain the pendulum parameters $\sigma$ and $\lambda$ and thus the corresponding spin $S=2\sigma\lambda$ as well as the coupling parameter $\gamma$, defined to be positive,
and we have $M_0=\int dv g_v$ and $M_1=\int dv v g_v$. In this way, four of the six parameters are given that determine a three-mode realization of the same pendular motion. The natural pendulum frequency (in units of $\mu$) is given by $\lambda^2=\wP^2+\Gamma^2$ and in our three-mode case $\lambda^2=v_1v_2v_3 M_0 M_1$, implying
$v_1v_2v_3=(\wP^2+\Gamma^2)/(M_0 M_1)$. On the l.h.s., $|v_1v_2v_3|<1$, suggesting that $\wP^2+\Gamma^2<|M_0 M_1|$. In our numerical examples this condition is certainly fulfilled, but it is not mathematically obvious if this is generally true
for any single-crossed spectrum that exhibits an instability. If it were not the case, a three-mode realization of the motion would not be possible.

Assuming this to be the case for a given $g_v$ we can choose the three-mode representation such that $v_1=-1$, $v_3=+1$, and $v_2=u$ with $-1<u<+1$. Then the equivalent three-mode system is given by
\begin{equation}
    S=2\wP
   \quad\hbox{and}\quad
    u=-\frac{\wP^2+\Gamma^2}{M_0 M_1}
\end{equation}
and
\begin{subequations}
\begin{eqnarray}
g_{v=-1}&=&\frac{S-M_1+u M_0}{2(1+u)}\ ,
\\
g_{v=u}&=&\frac{-S-u M_1+M_0}{1-u^2}\ ,
\\
g_{v=+1}&=&\frac{S+M_1-u M_0}{2(1-u)}\ .
\end{eqnarray}
\end{subequations}

In summary, we have found that three discrete modes behave like a stable or unstable flavor pendulum, the latter being described by only two parameters, the natural frequency $\lambda$ and spin $S$. Conversely, for a given pendulum with these parameters we can identify a two-parameter family of three-mode realizations.

\section{C.~Explicit solution for continuous spectrum}

If a single-crossed spectrum $g_v$ produces a coherent pendulum-like solution, we have seen that the motion of $\vect{D}_1(t)$ can be understood as a pendulum with parameters that can be extracted from $g_v$ without solving the EOMs. We have also seen that in this case the Bloch vectors $\vect{D}_v(t)$ are functions that one should be able to express as linear combinations of only three independent functions. We have seen in Supplement~B that we can identify three functions that we now call $\vect{P}_{v_i}(t)$ that reproduce the same pendulum with $\vect{D}_1(t)=\vect{P}_1(t)$. (In the corresponding discussion for slow modes these functions were called ``carrier modes'' \cite{Raffelt:2011yb}.) These three functions solve the EOM
\begin{equation}\label{eq:EOM-three}
    \dot{\vect{P}}_{v_i}=\mu\,v_i\vect{P}_{v_i}\times\vect{P}_1\ .
\end{equation}

We now transform the three functions to produce a continuous spectrum by virtue of
\begin{equation}
    \bar{\vect{P}}_v=\prod_{i=1}^3 (v_i-v)\,\sum_{i=1}^3\frac{v_i\vect{P}_{v_i}}{v_i-v}\ .
\end{equation}
The first factor was included to avoid a singularity when $v$ equals one of the discrete velocities. These new functions fulfill the original EOM
\begin{equation}
   \partial_t\bar{\vect{P}}_v=\mu\,v\bar{\vect{P}}_v\times\vect{P}_1
\end{equation}
as one can easily verify by inserting the definition of $\bar{\vect{P}}_v$ and using Eq.~\eqref{eq:EOM-three}. We may further define the unit vectors
\begin{equation}
    \vect{p}_v=\pm\frac{\bar{\vect{P}}_v}{|\bar{\vect{P}}_v|}
\end{equation}
with a possible sign change such that $p_v^z|_{t=0}=1$. Therefore, the solutions for the original modes are simply $\vect{D}_v(t)=g_v\vect{p}_v(t)$.

To summarize, if the spectrum $g_v$ reveals, in the linear regime, an instability we can construct the nonlinear solution 
for $\vect{D}_1(t)$ in the form of a pendulum, obtain three modes that produce the same pendulum motion, and construct the
explicit solution for $\vect{D}_v(t)$ and any other Bloch vector that follows the same EOM as $\vect{D}_v(t)$. In other words, the pendulum solution suggested by the information from the linear equations indeed fulfills the original EOMs also in the nonlinear regime.

\section{D.~Normal-mode analysis}

In the main text, we have briefly sketched the normal-mode analysis in the homogeneous case, leading to an eigenvalue equation that is very simple. However, it is also instructive to arrive at this result beginning with the inhomogeneous equations and taking the homogeneous limit in the end. The final result is the same, but it is nevertheless reassuring that there is no hidden issue of non-commuting limits. In principle, of course, this is just a step-by-step account of what can be found in the literature in various forms.

Our starting point is the two-flavor EOM, assuming axial symmetry, before taking the homogeneous limit:
\begin{equation}\label{eq:EOM-matrix}
 i\,(\partial_t+v\partial_z)\sD_v=\mu[\sD_0,\sD_v]-\mu[\sD_1,v\sD_v]\ .
\end{equation}
Here \smash{$\sD_0=\int dv\,\sD_v$} and $\sD_1=\int dv\,v\sD_v$ and we use the notation \smash{$\int dv=\int_{-1}^{+1}dv$} as in the main text. We here keep explicitly the first term on the r.h.s.\ without going to a co-rotating frame because we are interested in the real part of the dispersion relation that should be carefully distinguished from the overall precession caused by this neutrino-neutrino matter term which we follow carefully.

We recall that, in terms of the Bloch vector components, the lepton-number density matrices are
\begin{equation}
    \sD_v=\frac{1}{2}\begin{pmatrix}D_v^z&D_v^{xy}\\ D_v^{yx}&-D_v^z\end{pmatrix}
    =\frac{g_v}{2}\begin{pmatrix}s_v&S_v\\ S^*_v&-s_v\end{pmatrix},  
\end{equation}
where $D_v^{xy}=D_v^x+iD_v^y$ and the complex conjugate $D_v^{yx}=D_v^x-iD_v^y$. The ELN spectrum is the initial $g_v=D_v^z\vert_{t=0}$ and is assumed not to depend on space. In other words, we assume the initial setup to be homogeneous, but the solutions are allowed to be inhomogeneous. The diagonal and off-diagonal normalized components $s_v$ and $S_v$ follow our older notation and are not related to the particle-number matrices. 

The linear regime consists of the off-diagonal elements being small compared with the diagonal ones, in normalized form meaning that $|S_v|\ll1$ and the expansion is in powers of $S_v$. Taking the $z$-components at their initial value, the off-diagonal EOM is
\begin{equation}
    \bigl[i\,(\partial_t+v\partial_z)-(\Lambda_0-v \Lambda_1)\bigr]S_v=
    -\mu\!\int\!du\,g_u\bigl(S_u-vu S_u\bigr)\ ,
\end{equation}
where $\Lambda_0=\mu D_0$ and $\Lambda_1=\mu D_1=\mu D_1^z|_{t=0}$.

For a normal-mode analysis we seek plane-wave solutions of the form $S_v(t,z)=Q_v\,e^{-i(\Omega t-K z)}$, where $Q_v$ depends on the wave vector $(\Omega,K)$ and $K$ is the wavevector in the $z$ direction. The EOM in Fourier space thus is
\begin{equation}\label{eq:EOM-App}
    \bigl[(\underbrace{\Omega-\Lambda_0}_{\textstyle\omega})-v(\underbrace{K-\Lambda_1}_{\textstyle k})\bigr]Q_v=
    -\mu\!\int\!du\,g_u\bigl(Q_u-vu Q_u\bigr)\ .
\end{equation}
In the absence of interactions ($\mu=0$) the only solutions are $\omega=v k$, which are ``under the light cone'' defined by $\omega=k$, and have eigenfunctions $Q_v=\delta(\omega-v k)$. For nonzero $\mu$, these non-collective modes continue to exist with a more complicated singular $Q_v$
\cite{Capozzi:2019lso}.

In addition, collective modes appear which either have a real $\omega>k$ or a complex $\omega$ without restrictions on $k$. As a function of $v$, the r.h.s.\ of Eq.~\eqref{eq:EOM-App} has the form $a - b v$, where $a$ and $b$ are numbers that depend on the spectrum and on the solution, but not on $v$. Therefore, the eigenfunction is of the form
\begin{equation}
    Q_v=\frac{a-bv}{\omega-vk},
\end{equation}
implying
\begin{equation}
    a-bv=-\!\int\!du\,G_u\frac{a-bu-vu(a-bu)}{\omega-uk}\ ,
\end{equation}
where we have now absorbed $\mu$ in $G_v=\mu\,g_v$.

This equation must be true for all $v$, so we have two equations that can be written as
\begin{equation}\label{eq:Detab}
    \underbrace{\begin{pmatrix}I_0+1&-I_1\\  -I_1&I_2-1\end{pmatrix}}_{\textstyle \Pi} \begin{pmatrix}a\\ b\end{pmatrix}=0.
\end{equation}
Here the ``moments'' are
\begin{equation}\label{eq:moments}
    I_n(\omega,k)=\int du\,G_u\,\frac{u^n}{\omega-uk}.
\end{equation}
The dispersion relation follows from
\begin{equation}\label{eq:determinant}
    {\rm det}\,\Pi=(I_0+1)(I_2-1)-I_1^2=0.
\end{equation}
Once we have found $\omega(k)$ we can determine the eigenfunction up to an overall factor, i.e., for a chosen $a$ we can find $b$ or the other way around.

There is a nontrivial relation between the moments defined in Eq.~\eqref{eq:moments} as can be seen by the following manipulations:
\begin{equation}\label{eq:L0}
   \Lambda_0=\int du\,G_u=\int du\,G_u\frac{\omega-u k}{\omega-u k}=\omega I_0-k I_1
\end{equation}
and likewise
\begin{equation}\label{eq:L1}
   \Lambda_1=\int du\,G_u\,u=\omega I_1-k I_2.
\end{equation}
Therefore, two of $I_0$, $I_1$ and $I_2$ can be eliminated from the determinant condition Eq.~\eqref{eq:determinant} which thus can be written in three alternative forms in terms of only one of them. One case is
\begin{equation}
    I_0(\omega,k)=\frac{\Lambda_0(\omega+\Lambda_0)+k(k+\Lambda_1)}{\omega(\omega+\Lambda_0)-k(k+\Lambda_1)}\ .
\end{equation}
The physically homogeneous case $K=0$ implies that $k=-\Lambda_1$. Therefore, the determinant condition simplifies to
\begin{subequations}
\begin{equation}
    \omega I_0(\omega,-\Lambda_1)=\Lambda_0
\end{equation}
and with Eqs.~\eqref{eq:L0} and \eqref{eq:L1} implies
\begin{eqnarray}
   I_1(\omega,-\Lambda_1)&=&0,
   \\
   I_2(\omega,-\Lambda_1)&=&1.
\end{eqnarray}
\end{subequations}
If we insert these results in Eq.~\eqref{eq:Detab} we see that the second equation is fulfilled for any $a$ and $b$, whereas the first equation requires $a=0$. Therefore, we conclude that in the physically homogeneous case, the eigenfunction has $a=0$ and thus is proportional to $v$ with an arbitrary coefficient $b\not=0$. 

We thus recover the result derived in the main text where we started directly from the homogeneous EOM. In terms of physical variables, the eigenvalue is determined by $(\Omega-\Lambda_0)I_0(\Omega-\Lambda_0,-\Lambda_1)=\Lambda_0$. Here as always going to the co-moving frame in flavor space amounts to absorbing $\Lambda_0$ in $\Omega\to\omega=\Omega-\Lambda_0$ and not setting $\Lambda_0=0$.


\begin{figure}[]
\centering
\includegraphics[width=0.4\textwidth]{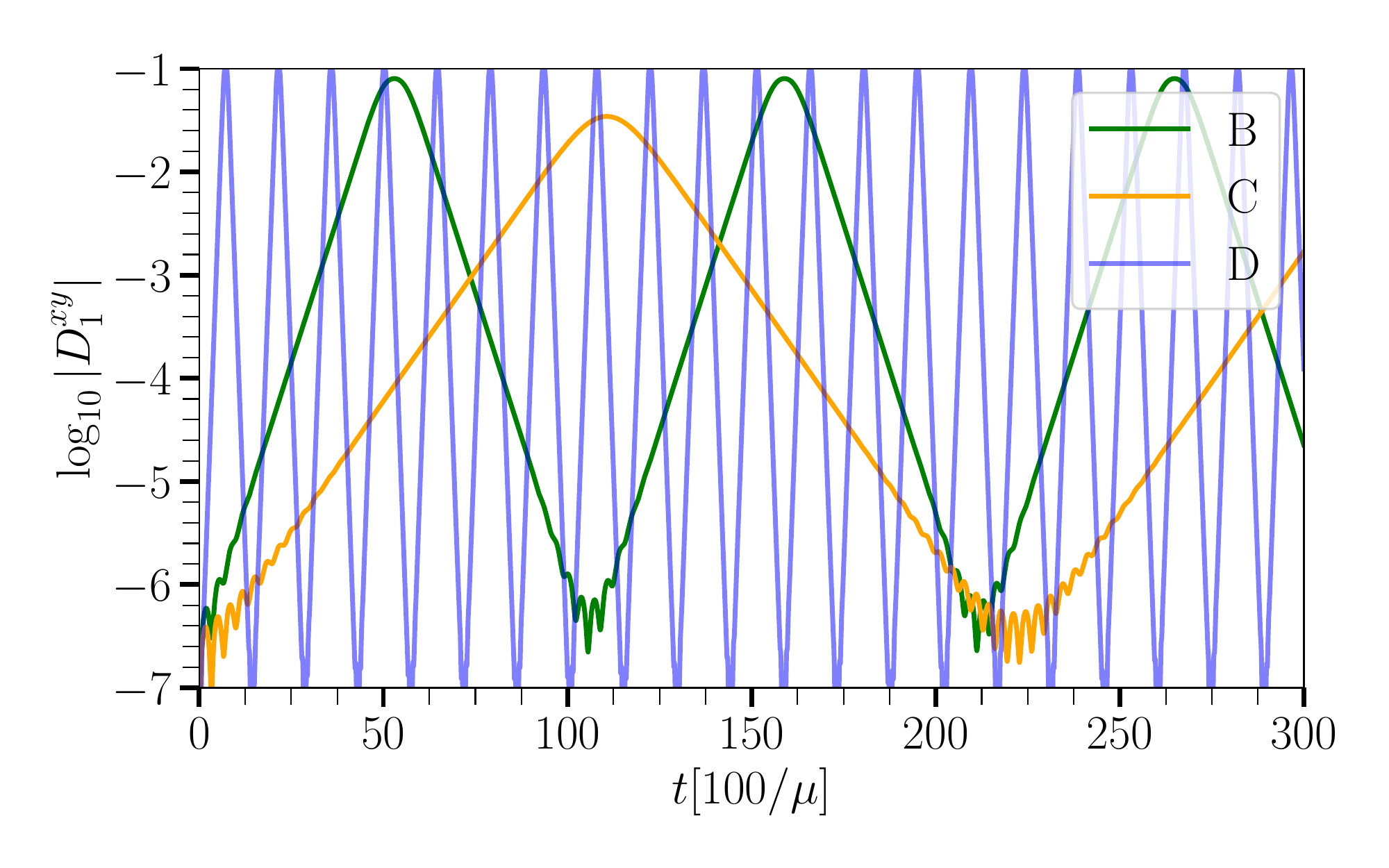}
\caption{Same as Fig.~\ref{fig:1b}, but for the evolution of the transverse components (flavor direction) of the lepton-number flux, $|D^{xy}_1| = \sqrt{ (D^x_1)^2 + (D^y_1)^2}$ for the unstable cases B--D. 
\label{fig:off-diag}}
\end{figure}

\section{E.~Further analysis of our numerical examples}
\begin{table}[ht]
 \caption{Solutions for the complex eigenfrequencies of our benchmark ELN configurations A--D (see main text).}
    \label{tab:solutions}
    \centering
    \begin{tabular*}{\columnwidth}{@{\extracolsep{\fill}}llllll}
    \hline\hline
     Case& $\Lambda_0$ & $\Lambda_1$ &$\omega_\pm=\wP\pm i\Gamma$&$\sigma$& $\cos{{\vartheta}_{\rm min}}$\\
         & [$\mu/100$] & [$\mu/100$] & [$\mu/100$]\\
     \hline
     A   & $-1.2666$ & $-4.2666$ & stable&---&---\\
     B   & $+0.7334$ & $-4.2666$ & $0.1828\pm0.1291\,i$ & $0.817$ & $+0.335$ \\
     C   & $+0.7388$ & $-3.2728$ & $0.2047\pm0.0584\,i$ & $0.962$ & $+0.849$ \\
     D   & $+4.7334$ & $-5.2665$ & $1.0743\pm1.1121\,i$ & $0.694$ & $-0.034$ \\
     \hline
    \end{tabular*}
    \vskip12pt
\end{table}

The solutions for the complex eigenfrequencies for our examples A--D (see main text) are summarized in Table~\ref{tab:solutions}. 
The analytical results are in excellent agreement with the numerical ones, as was already shown in Fig.~\ref{fig:1b} in the main text. In Fig.~\ref{fig:off-diag}, we show  the evolution of the $xy$ component that grows exponentially until the nonlinear regime is reached. The ``wiggles'' around the lowest points reflect the initial conditions (the small seeds) that excite all modes, but only the unstable one subsequently grows exponentially.

\begin{figure}[ht]
\centering
\includegraphics[width=0.4\textwidth]{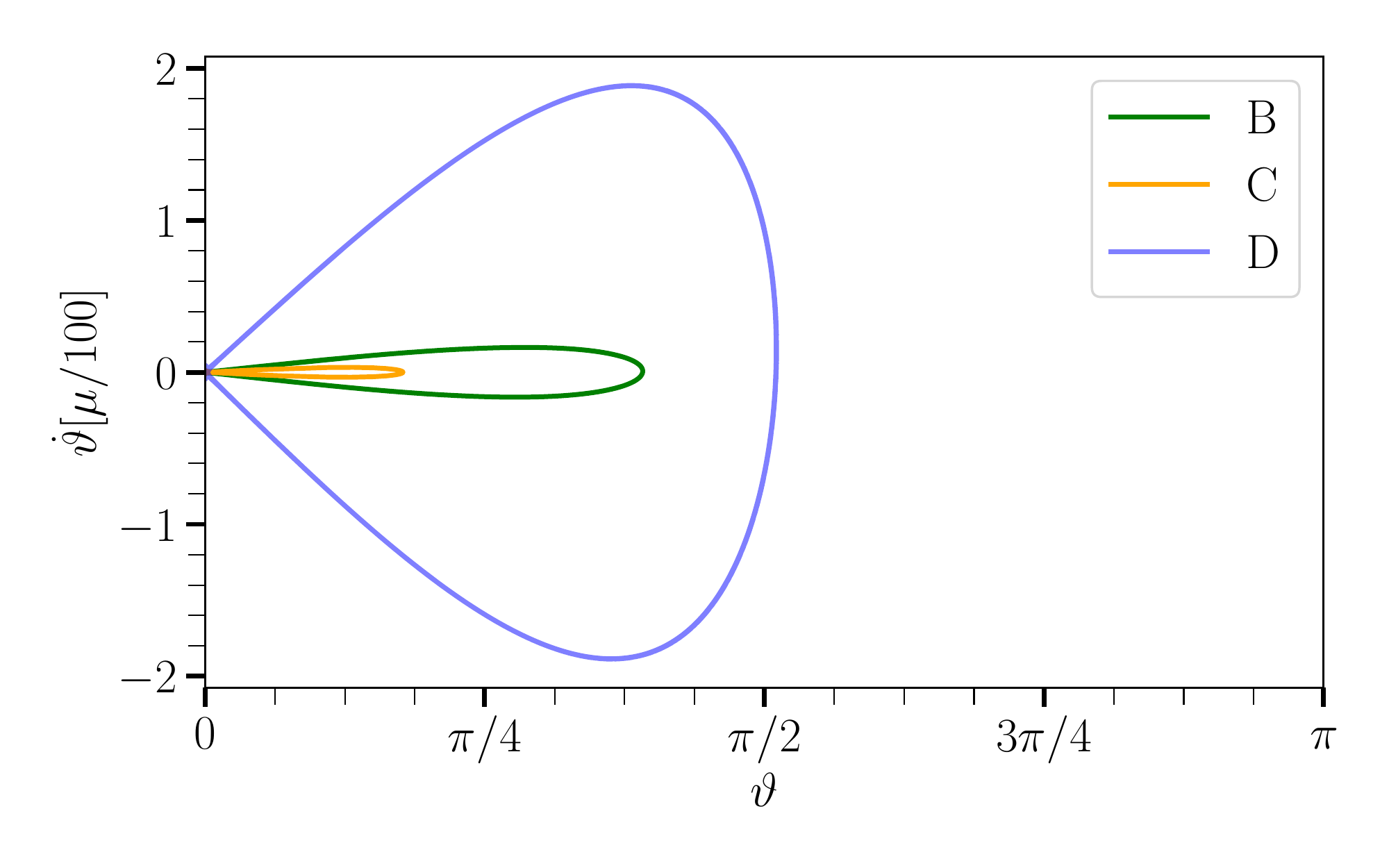}\\
\includegraphics[width=0.4\textwidth]{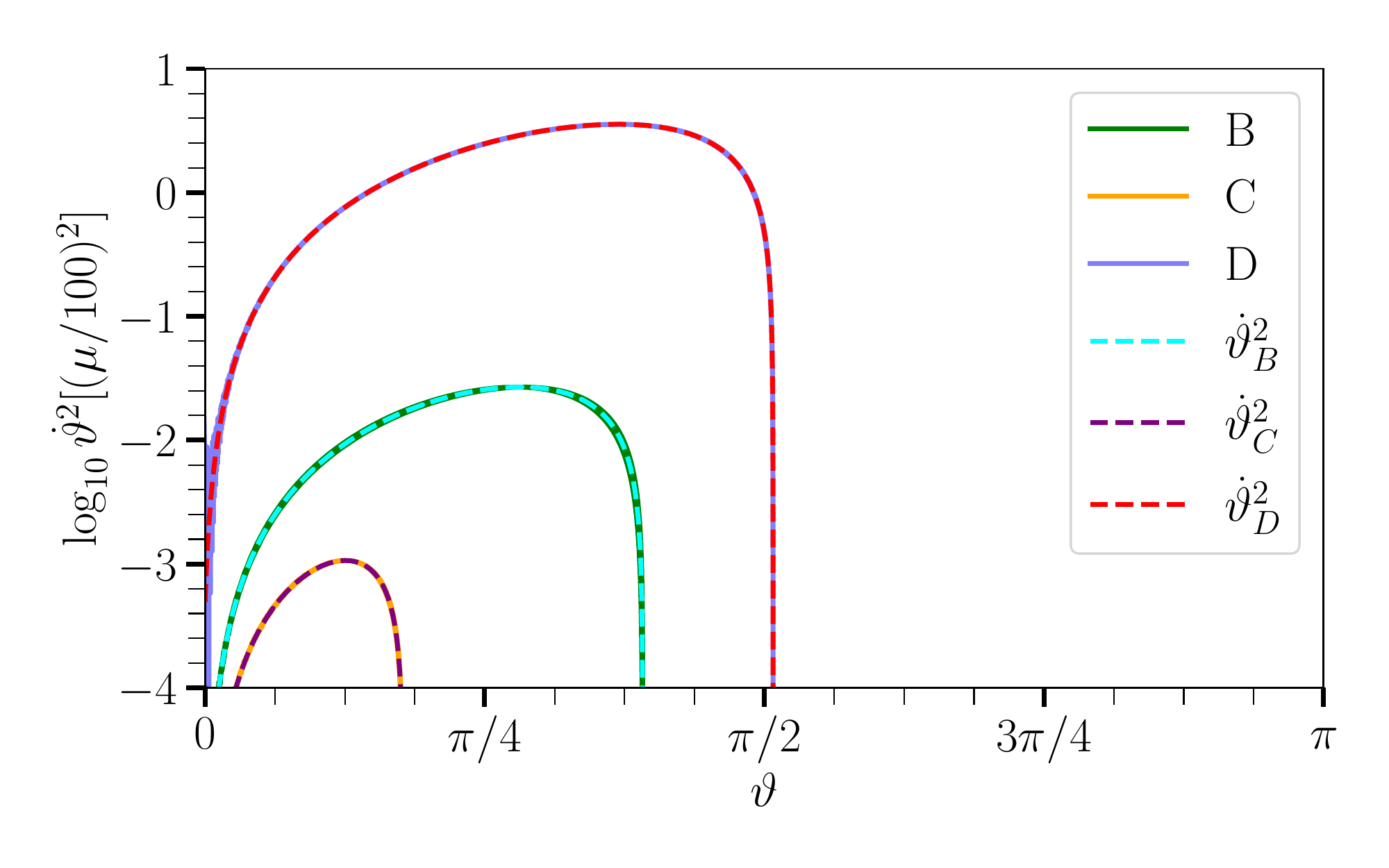}
\caption{Parametric plots of $\dot{\vartheta}$ (top)  and $\dot{\vartheta}^2$ (bottom) as  functions of $\vartheta$  for our unstable examples B--D. In order to favor a comparison, the bottom panel includes the numerical results for our three unstable examples B--D (solid lines) and the analytical ones (dashed lines). The agreement between the two is excellent.  \label{fig:thetaplot}}
\end{figure}

In order to highlight the periodic nature of the motion, we show in the upper panel of Fig.~\ref{fig:thetaplot} a phase diagram $(\dot\vartheta,\vartheta)$ derived from our solutions $\vartheta(t)$ and $\dot\vartheta(t)$. The motions continue to trace out their respective tracks.

To illustrate  the pendulum motion quantitatively, we show  $\dot{\vartheta}^2$ as a function of $\vartheta$ in the bottom panel, once more derived from the numerical solutions $\vartheta(t)$ and $\dot\vartheta(t)$. From Eq.~\eqref{eq:theta-eq} we glean that the motion is equivalent to a mass point with kinetic energy ${\dot\vartheta}^2$ that moves in a potential which is the negative of the r.h.s., so the numerically found ${\dot\vartheta}^2$ as a function of $\vartheta$ maps out the potential given on the r.h.s.\ of Eq.~\eqref{eq:theta-eq}. With the pendulum parameters shown in Table~\ref{tab:solutions}, the predicted curves are plotted as dashed lines, once more confirming the perfect agreement.


\end{document}